\pdfoutput=1

\documentclass[10pt]{article}
\usepackage{amssymb}

\usepackage{graphicx}
\usepackage{amsmath}

\newtheorem{theorem}{Theorem}
\newtheorem{acknowledgement}[theorem]{Acknowledgement}

\input{tcilatex}

\usepackage{pdfpages}

\begin{document}

\title{Cosmological Constant, Classical ''Vacuum'' and Special Relativity\\
(From the Lorentz boost to the Milgrom Acceleration)}
\author{Yves Pierseaux (ypiersea@ulb.ac.be)}
\maketitle

\begin{abstract}
\textbf{VERSION 1} \textbf{(results)} We show that Cosmological Constant $%
\Lambda $ is not optional in GR (\textit{general} relativity) because it is
required by SR (special relativity). This completely unexpected result is
obtained by introducing a minimal acceleration $\alpha =\sqrt{\Lambda }$
(Milgrom) in Einstein boost with Lorentz Transformation (LT). We prove that
hyperbolic rotation (LT) is an hyperbolic \textit{motion} with a centrifugal
acceleration. In Cosmological SR with $\Lambda $ (CSR or CR), the universe
is not only in expansion (with the law of Hubble) but even in accelerated
expansion (cosmological parameter $q=-1)$. In CR the constant $\Lambda $ is
naturally associated to the bending of light (Ishak and Rindler). Given that
the structure of space-time in Einstein's GR is determined by the \textit{%
presence} of matter and $\Lambda $ is associated to the \textit{absence} of
matter, we associate $\Lambda $ not to ''quantum'' vacuum but to classical
''vacuum'' of Minkowski's space-time. Finally we show that 1917 Einstein's
Cosmological Constant $\Lambda $ corresponds to 1906 Poincar\'{e}'s \textit{%
non-electromagnetic} negative pressure and we deduce density of
Poincar\'{e}'s relativistic fluid.

\textbf{VERSION 2 (methodology)}

In \textbf{\S 1} we switch the scale radius factor $a(t)$ of
Friedman-Lema\^{i}tre in the second member of Lorentz invariant $%
x^{2}-t^{2}=a^{2}$ $=H^{-2}=\alpha _{M}^{-2}=\Lambda ^{-1}$ ($c=1,H$ \ is
Hubble constant and $\alpha _{M}$ is Milgrom's acceleration).

In \textbf{\S 2} we sum up Einstein-Born-Rindler (EBR)'s model of
acceleration in SR. EBR's rigid motion is induced from non-relativistic 
\textit{rectilinear} uniformly accelerated motion.

In \textbf{\S 3} we propose another relativistic theory of acceleration
induced from non-relativistic uniformly \textit{circular} accelerated motion
given that Euclidean rotation (centripetal acceleration) must be replaced by
Hyperbolic Rotation \textit{motion} (\textit{centrifugal} acceleration).
This new theory of acceleration in Lorentz boost involves not a rigid motion
but an elastic expanding motion. We introduce a \textit{new symmetry} in
Minkowski's basic diagram of (scale) hyperbolas by adopting a new definition
of invariant proper distance in such a way that the LTed proper distance be
a dilated distance (symmetrically with invariant proper time). By giving up
Einstein's definition of length contraction, Minkowski's calibration
hyperbola (Passive LT) and Born's acceleration hyperbola (Active LT) become
identical.

In \textbf{\S 4} we introduce a \textit{basic minimal acceleration}
(Milgrom) in Lorentz boost with fundamental emission of radiation (CBR). By
renormalization of Minkowski's metric, we show that classical ''vacuum'' is
perfectly compatible with an expanding universe (Bondi's factor $k$) where
the Hubble radius $a=R_{H}$ is an hyperbolic horizon. CR is distinct from
the ''Doubly Special Relativity'' (DSR) approach, in that it does not look
for a new invariance group adapted to Einstein's definition of distance, but
for a new definition of distance, adapted to the LT.

\textbf{FIGURES} We suggest a geometrical version of this paper with \textbf{%
5 figures}. The principle of correspondence circle-hyperbola (PCCH) is
deeply inscribed in Hyperbolic rotation (LT). This principle is valid for
physical uniformaly accelerated \textit{circular} motion $a=\frac{v^{2}}{r}$
and unfformly accelerated \textit{hyperbolic} motion $\alpha =\frac{c^{2}}{X}
$. Both motion have a center but if the first one is centripetal, the second
one is \textbf{CENTRIFUGAL}.
\end{abstract}

\section{Cosmological constant as second invariant in CSR (or CR)}

\bigskip\ Let us consider fundamental hyperbolas along $Ot$ and $Ox$ in
Minkowski's space-time (at one space dimension with the axis, $x,$ $t$ of
system K with light velocity $c=1)$. The two hyperbolas determining the
units of measure ($x^{2}-t^{2}=\pm 1)$ are called hyperbolas of scale or
calibration (\cite{1}). We focus the attention on the along $Ox$ hyperbola
defined with invariance of \textit{space interval} by a \textit{passive}
Lorentz Transformation ($PLT$) (5) $x^{\prime }=OP^{\prime }=O^{\prime
}P^{\prime }$ $(t^{\prime }=0)$ (\textbf{Fig1}, light asymptotes and
standard representation of primed axis $x^{\prime },$ $t^{\prime }$ ''in
scissors'' or \textit{hyperbolic rotation} of system K')

\begin{equation}
x^{2}-t^{2}=x^{\prime 2}\ (t^{\prime }=0,\text{calibration hyperbola})\qquad
\qquad x^{2}-t^{2}=0\ \text{(standard metric asymptotes)}
\label{1-asymptote}
\end{equation}
Given that $x^{\prime }$ (the hyperbolic radius of curvature) can be as
large as we wish, the calibration hyperbolas disappear at the infinity and
we have only one invariant $c=1$ in standard configuration. Let us however
note that, only the \textit{finite} interval involves, according to
Minkowski, that \textit{''space by itself and time by itself are doomed to
fade away into mere shadows, and only a kind of union of the two will
preserve an independent reality''}. Indeed an ''infinite'' interval ($%
x^{2}-t^{2}$ \ \ $=$ \ $\infty $, see 21) should mean that the independent
space is given for any $t$ and therefore the return of the shadow (Absolute
Space $Ox$, $\forall t$). In order to stay in relativistic configuration
suppose now that we have a very small but non nul constant $\Lambda \neq 0$
in such a way that

\begin{equation}
R_{C}^{2}=x^{\prime 2}=x^{2}-t^{2}\nrightarrow \infty \qquad \qquad
x^{2}-t^{2}=\Lambda ^{-1}\text{ \ \ \ \ \ \ \ \ \ }c=1
\label{2-finite interval}
\end{equation}
Such a reformed Minkowski's metric no longer is flat because we have an
hyperbolic global curvature $\varrho _{h}=\frac{1}{R_{c}}(\Lambda =\frac{1}{%
R_{c}^{2}})$ \ In this case we have a constant radius $R_{C}$ of the
universe and therefore a static universe incompatible with the observed
universe ''in dilation''. However, by the same argument with the other
hyperbola ($Ot$), we could claim that we should have an observed dilated
time (in K, \textbf{Fig1}) of Universe (see \S 3.4).

Let us now underline that the right branch $x>0$ of $Ox$ hyperbola $%
(t^{\prime }=0)$ can also represent a \textit{worldline} of an uniformly ($%
\alpha )$ accelerated particle $P^{\prime }.$ According to Rindler (\textbf{%
Fig2}, \cite{2a})

\begin{quotation}
Consider a rod of arbitrary length resting along the x axis of Minkowski
space. A time $t=0$ we wish to give one point of the rod a certain constant
proper acceleration and we want the rod as a whole to move rigidly, i.e. in
such a way that the proper length of each its infinitesimal elements is
preserved. It turns out that each point of the rod must then move with a
different though also constant proper acceleration, the necessary
acceleration increasing in the negative direction and become infinite at a
well-defined point of the rod; the rod can evidently not to be extended
beyond or even quite up to that point, since an infinite proper acceleration
corresponds to motion at the speed of light. If we arrange things so that
this cutoff point lies originally at the origin the equation of motion of
the point originally at $x^{\prime }=X$ $\ $is (2) \ We take X as a
convenient spatial coordinate on the rod.
\end{quotation}

\begin{equation}
x^{2}-t^{2}=X^{2}=\alpha ^{-2}\text{ }(t^{\prime }=0,\text{ }X>0)\qquad
\qquad x^{2}-t^{2}=0\text{ \ \ \ (}x>0)  \label{3-acceleration hyperbola}
\end{equation}
(end of quotation). Born's rigid motion suppose a set of hyperbolic
worldlines of points belonging to a rigid rod $x^{\prime }=O^{\prime
}P^{\prime }=X.$ This basic motion with constant interval, is induced by
successive $ALT$ (Active Lorentz transformation, \S 2.2) which boosts the
particle fixed in $K^{\prime }$ (the rigid rod) to a larger velocity. Given
that in standard configuration the proper acceleration $\alpha $ can be as
small as we wish $\alpha \rightarrow 0$ and so the hyperbola disappears at
the infinity\footnote{%
We underline that in cosmology Minkowski's metric is characterized by the
disappearance of Minkowki's calibration hyperbolas. Only the asymptotes
survive to the extinction (only one invariant). So Minkowski's metric is a
non-calibrated metric (see equation 21 renormalized by 34).}: it only
remains the light cone that defines the standard Minkowski's metric with
only one invariant With $a=\gamma ^{3}\alpha =0$ (14)$,$ the geodesics are 
\textit{straight lines} in \textit{flat standard }Minkowski's metric However
in special relativity (\S 3-2) the acceleration is a \textit{spacelike}
four-vector whose norm must necessarily be larger than zero $\alpha >0$.
Suppose now the existence\ of a minimal acceleration $\alpha _{M}$ (3) which
corresponds to the relativistic inexistence of an infinite interval (2).

\begin{equation}
\text{(left side) \ \ \ \ \ \ }\frac{1}{R_{H}^{2}}=\varrho ^{2}=\text{ \ \ }%
\mathbf{\Lambda }\text{ \ \ \ }=\text{\ }\alpha _{M}^{2}=H^{2}\qquad \text{%
(right side)}\qquad (c=1)  \label{4-cosmological constants}
\end{equation}

where $R_{H}$ is the observable radius of Hubble and $\alpha _{M}$ is the
observable acceleration of Milgrom (\cite{3a}). We rediscover the empirical
standard values of constants $x^{2}-c^{2}t^{2}=\frac{c^{4}}{\alpha ^{2}}%
=R_{H}^{2}$.$(c=3.10^{8}m/s$, $T_{H}\approx 3.10^{17}s,$ $\ \alpha
_{M}=10^{-9}m/s^{2}$ \cite{3b})$.$ The cosmological constant ($\Lambda =%
\frac{1}{R_{H}^{2}}$, standard value defined by static 1917 Einstein's
model) is a second invariant in our new SR which connects two previously
independent empirical quantities $R_{H}=cT_{H}$ and \ $\alpha _{M}$. This
new completed SR that we suggest to call \textit{''Cosmological (special)
Relativity''} (\textbf{CR}) introduces in physics a \textit{scalar} field of
acceleration\textit{\ }$\alpha _{M}$ that is non-gravitionnal because the
fundamental acceleration depends on the distance and not on the square of
the distance.

The constant $\Lambda $ can be see either from the left side or the right
side of equations (4).

In the first case (left side) we have a geometrical interpretation: the
geodesic is no longer a straight space-time world line but an hyperbolic
spacetime worldline that can be connected with hyperbolic trajectories of
Pioneer sounds (\cite{4}).\ Exactly as in Euclidean geometry $R^{-1}$
represents the global constant curvature of circle, $\varrho =$ $R_{H}^{-1}$
represents in Minkowskian geometry the \textit{global} constant curvature of
the hyperbola. But the difference is that the radius $r$ of the circle can
be as great as we wish $r\longrightarrow \infty $ (globally the curvature is
zero) the maximal interval-distance is $R_{H}$ in CR (globally the curvature
is non-zero) \ The basic heuristic in CR is based on the relativistic
transformation of circle into hyperbola (that is well known, \textbf{Fig1})
and therefore also on the relativistic transformation of non-relativistic
circular uniformly accelerated \textit{motion} into hyperbolic rectilinear
uniformly accelerated motion (that is not very well known, \textbf{Fig3}).
This ''principle of correspondence ''circle-hyperbola'' \textbf{(PCCH)}
plays a basic role in our analysis (5 Figures).

In the second case (right side) we\ have not only a cosmological
interpretation of Milgrom's acceleration $\Lambda =\alpha _{M}^{2}$ (\cite{5}%
) but also, very curiously, Hubble physical constant $H$ $(c=1)=\alpha _{M}$%
. The presence of this constant of expansion in (4) and therefore in (3 \&
2) is very strange because Minkowski's space-time is renowned incompatible
with expanding universe.\ The fact that the Hubble radius is a constant ($%
R_{H}=R_{C})$ is not in contradiction with the observed expansion of the
universe because we will prove that this radius is an \textbf{hyperbolic
horizon} which involves the law of Hubble (37).\ \ 

However there is a serious problem in $CR$ because our theory is a purely
logical building based on the identification (Fig1 \& Fig2) between
Minkowski's hyperbola ($PLT$ without explicit acceleration) and Born's
hyperbola ($ALT$, \ with an acceleration). This identification first
involves the coincidence with $O$ and $O^{\prime }$, center of rotation in
both cases in CR (\S 3) does not correspond to the most sophisticated model
of rigid motion (\S 2, according to Rindler the ''end ($O^{\prime }$) of rod
is a photon''). Rindler's representation of Einstein's contraction involves
that $O^{\prime }$ moves on the asymptote (its own infinite hyperbola) with
respect to $O$. This is the reason why the two hyperbolas are generally not
identified. Our identification involves a physical synthesis, based on 
\textbf{hyperbolic rotation motion}, between $PLT$ and $ALT$ and therefore
the determination of a \textbf{fundamental acceleration in Lorentz boost }%
(\S 4).

\section{\protect\bigskip Einstein-Born-Rindler's rigid motion and Cosmology}

Einstein introduced his theory of contraction of length in 1905 (second
paragraph) by using rigid rods (\cite{7}). Born defined in 1909 the \textit{%
rigid motion} of each points of Einstein's rigid rod, by using accelerated
hyperbolic motion of a particle (\cite{8}). Rindler found in 1966 an
original metric in cosmology on the basis of Einstein-Born's rigid motion in
special relativity (\cite{2c}).

\subsection{Einstein's contraction of rigid rod by $PLT$}

Einstein considers two systems (rigid rods), $K$ and $K^{\prime }$, in
standard configuration. Lorentz Transformation (LT), at one space dimension
is: 
\begin{equation}
x^{\prime }=\gamma (x-\beta t)\ \qquad t^{\prime }=\gamma (t-\beta x)\qquad 
\text{(a)}\qquad x=\gamma (x^{\prime }+\beta t^{\prime })\ \qquad t=\gamma
(t^{\prime }+\beta x^{\prime })\qquad \text{(b)}  \label{5-LT}
\end{equation}
with instantaneous coincidence of $O\equiv O^{\prime }$ at $t=t^{\prime }=0$
and space-time $c=1$ units $"c=1"$, $\gamma =(1-\beta ^{2})^{-\frac{1}{2}}$.
In Einstein's paragraph 2 (\cite{7}), the system $K^{\prime }(x^{\prime
}=O^{\prime }P^{\prime }=L)$ is in uniform translation with respect to the
system $K$ for \textit{any} $t$ ($t<0$ and $t\geq 0).$ Einstein uses \textit{%
a passive LT} in order to define the length in K of a moving rigid rod $%
O^{\prime }P^{\prime }$ of $K^{\prime }$ with the simultaneity of two events 
$\ t=0$ in K (the difference of space coordinate of two ends $O^{\prime }$
and $P^{\prime }$ of the rigid rod at the same time in $K$). By passive
using of first $LT$ (5a), we find the coordinates in $K$ of $O^{\prime }$
and $P^{\prime }$, respectively for finite (6a) and differential length (6b):

\begin{eqnarray}
\ t &=&0\Rightarrow x=\gamma ^{-1}x^{\prime }\text{\ \ \ \ \ \ \ \ \ \ \ \ (}%
x^{2}=x^{\prime 2}-t^{\prime 2}\text{)\ \ \ \ \ \ \ global \ (a) \ }
\label{6-Einstein's contraction} \\
\text{ \ }dt &=&0\Rightarrow \text{\ }dx=\gamma ^{-1}dx^{\prime }\text{ \ \
\ \ (}dx^{2}=dx^{\prime 2}-dt^{\prime 2})\text{\ \ \ \ local\ \ (b)}  \notag
\end{eqnarray}

We note that Einstein's contraction suppose $t=0$ and not $t^{\prime }=0$.

\subsection{Born's rectilinear uniformly accelerated rigid motion by $ALT$}

\bigskip Let us summarize briefly Born's procedure. In SR, the acceleration
is not an invariant ($a^{\prime }$ and $a)$ and only a \textit{proper}
constant acceleration in K' $\alpha _{\text{Born}}=$ $a^{\prime }$ can be
defined at the\ condition to do ''$v^{\prime }=0"$ in the components of
acceleration four-vector $A^{\prime }:$ $\gamma ^{\prime }(\gamma ^{\prime
}a_{x}^{\prime }$ $+$ $\frac{d\gamma ^{\prime }}{dt^{\prime }}v^{\prime }%
\mathbf{,}$ $\gamma ^{\prime 3}v^{\prime }.\frac{dv^{\prime }}{dt^{\prime }}%
c)$ (at one space dimension): 
\begin{equation}
v^{\prime }=\frac{dx^{\prime }}{dt^{\prime }}=0\ \ \text{and}\ \ \gamma
^{\prime }(v^{\prime })=1\ \ \Longrightarrow \ \ \ \ \alpha =a^{\prime }=%
\frac{dv^{\prime }}{dt^{\prime }}=\gamma ^{3}(v)\frac{dv}{dt}=\frac{%
d(v\gamma )}{dt}  \label{7-differential equation}
\end{equation}
The comoving system with the particle $P^{\prime }$ is thus at each instant
a new system, $K^{\prime }$, defined by successive $ALT$ with ''$v^{\prime
}=0$ but $dv^{\prime }\neq 0$'' ($dv^{\prime }=\gamma ^{2}.dv$) and with ''$%
t^{\prime }=0$ but $dt^{\prime }\neq 0"$. In each K' system, the equation of
the $O^{\prime }x^{\prime }$ axis is $t^{\prime }=0$. The origin $O^{\prime
} $ of successive $K^{\prime }$ is generally located in $P^{\prime }$(Born);
we will adopt the standard configuration.(Rindler), by locating the particle 
$P^{\prime }$ at a distance $x^{\prime }=X$ (Born-Rindler's notations,\cite
{2b}) from $O^{\prime }$ in K' at the coordinate $x_{0}=X$ in $K$ ($%
t=t^{\prime }=0)$. The differential equation $(c=1)$ is $\alpha dt=$ $%
d(\gamma \beta )$. A first integration is carried out with $\beta \ =0$ at $%
t=0$: $\alpha t=\gamma \beta $ or:\ 
\begin{equation}
\alpha dt=d(\gamma \beta )\qquad \Longrightarrow \qquad \beta =\frac{dx}{dt}=%
\frac{\alpha t}{\sqrt{1+\alpha ^{2}t^{2}}}  \label{8-first integration}
\end{equation}
A second integration yields $x=\frac{1}{\alpha }\sqrt{1+\alpha ^{2}t^{2}}$ \
with $x_{0}=$ $\frac{1}{\alpha }$.and therefore the right branch of
hyperbola:

\begin{equation}
x^{2}-t^{2}=x^{\prime 2}=\alpha ^{-2}\ =X^{2}\ \ (X\geq 0)\text{\ }\qquad 
\text{\ (a)}\ \ \ x=\gamma X\text{ \ \ \ (b) \ \ \ \ \ \ \ \ }\ t^{\prime
}=t-\beta x=0\text{ (c) }  \label{9-second integration}
\end{equation}
The accelerated hyperbolic motion is resolved by successive $ALT$ into an
infinity of instantaneous inertial motions \textbf{(Fig2)}. This is Born's
definition of \textit{rigid motion} of the rod: at each instant $t^{\prime
}=0$, the entire rod (and the infinitesimal rigid rod $dX=dx^{\prime }$) is
at rest in its proper system\ K'(in Born-Rindler's notations $X=\alpha
^{-1}) $. It is then impossible to separate the motion of the particle $%
P^{\prime }$ from the motion of the system $K^{\prime }$(the rigid rod $%
O^{\prime }P^{\prime }$) that drags it along. In other words, it is
impossible to separate successive $ALT$ (succession of events) from
successive boosts of systems $K^{\prime }$.

EBR's rigid motion is a relativistic interpretation of \textit{rectilinear
non-relativistic accelerated} motion which involves that the rod $O^{\prime
}P^{\prime }$ is in motion with respect to the origin $O$. So in Rindler's
diagram $(x,t)$ the end of the rod $O^{\prime }$ for $t>0$ is in motion on
its own infinite hyperbola (\textbf{Fig2}), i.e. the asymptote.We underline
that EBR's rigid motion is not correlated (PCCH, \S 1) with a rotation
because in any rotation (including hyperbolic rotation) we have a center or
a fixed point ($O^{\prime }\equiv O)$. This is the reason why Rindler adopts
a basic equation which result from differentiating for constant $t,$ $%
xdx=XdX,$ introducing Einstein's contraction in rigid motion. 
\begin{equation}
\text{\ }x=\gamma X\text{ \ (''dilation'' by }ALT\text{)}\qquad \qquad 
\overset{xdx\text{ \ }=\text{ \ }XdX\text{ \ ????}}{\Longleftrightarrow }%
\qquad dx=\gamma ^{-1}dX\text{ (contraction by }PLT\text{)}
\label{10-contraction}
\end{equation}
K'. The problem in Einstein-Born's theory is that the entire rod $O^{\prime
}P^{\prime }$ is defined at constant $t^{\prime }=0$ (simultaneity) and
Einstein's contraction supposes also ''at constant $t"$ (also simultaneity,
Rindler's representation with $O^{\prime }P^{\prime }$ \textbf{Fig2}). The
quadratic form (6a) is not the quadratic form (9a): there is an asymmetry (%
\textbf{Fig1}). We underline that in Born's theory the infinitesimal
''proper length'' $dX=dx^{\prime }$ ($dt^{\prime }=0$) appears ''as an
invariant'' exactly as Minkowski's proper time $dt^{\prime }=d\tau $ ($%
dx^{\prime }=0$) appears as an invariant (see note 3). But in standard SR
this analogy must be limited because if the space interval defines a
distance exactly as the time interval defines a duration, we obtain a
dilation (9b) in both cases.

\subsection{Rindler's time parameter on rigid rod and cosmological
non-Minkowskian metric}

\bigskip Rindler gives in 1966 a cosmological interpretation of Born's
theory. He focuses at first the attention on the fact that there is in each
point of the rod a ''a different though also constant proper acceleration''
but also on a singularity because the rigid rod (the ''proper length'')
''ends in a photon'' (infinite acceleration $\alpha =\infty $). In order to
develop analogy with Schwarschild metric, Rindler then introduces a
parameter $T$ \ for observers with clocks on the rod (given that $t^{\prime
}=0$):

\begin{quotation}
It can be seen that the proper acceleration of the point $X$ of the rode is $%
1/X$. \ Hence an observer at $X$ feels a constant gravitational field of
intensity $1/X.$ The observers on the rod can so synchronize their clocks
that each sees all the other clocks neither gain nor lose relative to his
own; each observer \ must simply speed-up his proper clock by a factor equal
to the reciprocal of his coordinate X. Let $T$ denote this new time.
\end{quotation}

With Rindler's time parameter $T=\beta _{H},$ the hyperbolic velocity and
not the proper time $\beta _{H}=\alpha \tau $), one obtains a
non-Minkowskian metric 
\begin{equation}
ds^{2}=dx^{2}-dt^{2}=dX^{2}-X^{2}dT^{2}\ =dX^{2}-X^{2}d\beta _{H}^{2}\ \ \ \ 
\text{\ (a)\ \ }\ T=\alpha \tau \ \ \ \rightarrow \ \ \text{signature}\ (1,%
\text{ \ }-X^{2})\ \ \ \ \ \ \ \text{(b)}  \label{11-Rindller's metric}
\end{equation}

Ellis and Williams underline that Rindler's metric ($1,$ $-X^{2})$ is based
on ''the boost-invariance of the proper length of the rod'' or
''boost-invariance of flat space-time''(\cite{9})$.$ The fact that the
''boost-invariance'' involves a non-Minkowskian space-time is not consistent
with the fact that the boosts in question are... Lorentz boosts. We need
therefore a complete theory of $ALT$ or Lorentz boosts (\textit{hyperbolic
rotations}). Rindler finally does not use the parameter $\tau $ because the
proper time $\tau =$ $\beta _{H}X$, in a state of acceleration of \ $%
K^{\prime }$ systems, is a different one in each point of the rod\footnote{%
Let us however think about the following ''Gedanken Experiment''. Suppose a
rigid rod $O^{\prime }P^{\prime }$ in prerelativistic kinematics at rest in
the system K. If we push in $t^{\prime }=0$ the rod with a finite constant
acceleration in O', instantaneously P' is in moving with the same finite
constant acceleration. The two ends $O^{\prime }$ and $P^{\prime }$ of the
rods are instantaneously boosted at the velocity $\beta $ and therefore at
the same\ Rindler's time $T_{O^{\prime }}=T_{P^{\prime }}.$ This
instantaneity\ at distance however cannot be true in a completely
relativistic kinematics.$\ \ \ $}$\ \ \ \ \ \ \ \ \ \ \ \ \ \ \ \ \ \ \ \ \
\ \ \ \ \ \ \ \ \ \ \ \ \ \ \ \ \ \ \ \ \ \ \ \ \ \ \ \ \ \ \ \ \ \ \ \ \ \
\ \ \ \ \ \ \ \ \ \ \ \ \ \ \ \ \ \ \ \ \ \ \ \ \ \ \ \ \ $\ \ \ \ 

\section{Theory of acceleration in Lorentz boost (Hyperbolic ROTATION\
Motion and synthesis PLT-ALT)}

EBR refers to Einstein's contraction ($PLT$, his paragraph 2) but not
directly to Einstein's boost (his paragraph 3). \ In other words EBR's
demi-hyperbola \textbf{(Fig2)} is not concentrated on the transition from $0$%
-velocity to $\beta $-velocity ($ALT)$ \textbf{(Fig3)}$.$ This is a natural
interpretation of Born's integral hyperbola which is deduced from a
differential equation (8a) $\alpha dt=$ $d(\gamma \beta )$ (unlike
Minkowski's hyperbola). $\ $We now focus the attention on the element of 
\textit{arc of hyperbola} $P_{0}^{\prime }P^{\prime }$ \textbf{(Fig3)}, i.e.
on Einstein's boost.\ 

Einstein considers two systems K and K', at rest relatively with each other.
He places a rigid rod in K, of length $L=OP$, and an identical rigid rod in
K', of identical length $L=O^{\prime }P^{\prime }$ when both systems are at
rest. Einstein then ''boosts'' the K' system at $t=t^{\prime }=0$, to bring
it to a cruise speed of $\beta $ ; this calls for an acceleration. Einstein
explicitly states (in 1907) that the eventual consequences of this
acceleration $\alpha _{BOOST}$ on the $O^{\prime }P^{\prime }$ rod disappear
in K' as soon as the velocity becomes uniform (a statement we do not
question). Einstein\ however does not say anything about this acceleration
and how to compute this acceleration $\alpha $. Let us now develop into 
\textit{three successive phases} (\textbf{Fig3)} a theory of acceleration in
Einstein's boost with $ALT$ $(0\rightarrow \beta )$.

PHASE 1 \qquad\ The rod first is at rest $(v=\beta =0)$ for $t<0$

PHASE 2 \ \qquad Then in $t=t^{\prime }=0$ the rod is submitted in each
point to a constant acceleration $\alpha $ (\textit{arc of hyperbolic
rotation }$P_{0}^{\prime }P^{\prime }$).

\qquad \qquad \qquad\ \ \ \ The element of angle that subtends the element
of hyperbolic arc is hyperbolic velocity $d\beta _{H}=\alpha d\tau $\qquad

PHASE 3 \qquad\ Finally at $\ t=\beta \gamma X>0,$ the rod reaches its
cruise speed of $\beta $ (\textit{tangent }in $P^{\prime }$, $M^{\prime }$
is the intersection with asymptote)$.$

The end of the rod, the particle $P^{\prime }$ (the end of the rod) is
transported by an active LT from the point $P_{0}^{\prime }$ $(0,$ $X)$ to
the point $P^{\prime }(\gamma X,$ $\gamma \beta X)$ in K i.e. 
\begin{equation}
x^{2}-t^{2}=\alpha ^{-2}\ =X^{2}\ \ (\text{ }t\geq 0)\text{ \ \ \ \ \ \ }P(0)%
\overset{\text{\textbf{hyperbolic arc}}}{\rightarrow }P(\beta )\ \ \ \ \ \
\alpha _{\text{Boost}}(P^{\prime })=\alpha _{\text{Born}}\text{ (}ALT=\alpha
LT)\   \label{12-Lorentz boost}
\end{equation}
Given that, in hyperbolic motion the particle $P^{\prime }$ and its proper
comoving system $K^{\prime }(O^{\prime })$ are completely inseparable, $%
\alpha \ $is therefore the acceleration in any Einstein's boost $%
0\rightarrow $ $\beta $. Our new theory of Einstein's boost of the rod $%
O^{\prime }P^{\prime }$ is exactly Born's hyperbolic motion of particle $%
P^{\prime }$ if and only if we admit that $O^{\prime }$\textit{\ coincide
with }$O$\textit{\ }at $t=t^{\prime }=0,$\textit{\ for any velocity}: $%
\forall \beta $ \textbf{(}this is not true with Rindler's representation of
contraction, see the contrast \textbf{Fig1\&2)}$.$

\subsection{Hyperbolic Rotation motion with \textit{centri}fugal (expanding)
acceleration}

Our theory of Einstein's boost suppose that $ALT$ must be consistent with $%
PLT$. Hyperbolic Rotation with $t^{\prime }=0$ is:$\ \ \ \ $ 
\begin{eqnarray}
x^{\prime } &=&x\cosh \beta _{H}-t\sinh \beta _{H}\qquad \ \ t^{\prime
}=t\cosh \beta _{H}-x\sinh \beta _{H}\text{ \ \ \ (a) \ \ \ }  \TCItag{5-HR}
\\
\text{\ \ }x &=&x^{\prime }\cosh \beta _{H}+t^{\prime }\sinh \beta _{H}\text{
\ \ \ \ \ }t=t^{\prime }\cosh \beta _{H}+x^{\prime }\sinh \beta _{H}\text{ \
\ \ (b)}  \notag \\
\text{ \ \ \ }x^{\prime } &=&x\cosh \beta _{H}-t\sinh \beta _{H}=\alpha
^{-1}\qquad t^{\prime }=0\text{ \ \ \ \ (a)}\ \ \ \   \notag \\
\qquad \text{\ }x &=&x^{\prime }\cosh \beta _{H}=\gamma \alpha ^{-1}\qquad \
\ \ \ t=x^{\prime }\sinh \beta _{H}=t=\beta \gamma \alpha ^{-1}\text{\ \ (b)}
\notag
\end{eqnarray}
In order to obtain this fundamental coherence (\textbf{Fig1} \& \textbf{Fig3}%
), we have to admit that $O^{\prime }$($\equiv O)$ (with $\alpha =\infty $)
reaches instantaneously the $\beta -($or $\beta _{H}$)$-$velocity \ (the
center of hyperbolic rotation is the same): 
\begin{eqnarray}
&&\ \ \ \ \ \ P^{\prime }(X)\overset{\text{ Hyperbolic rotation (ALT)}}{%
\rightarrow }P^{\prime }(\gamma X)\text{\ \ (a)}  \notag \\
O &\equiv &O^{\prime }\ \ (\forall \beta ,\text{ }\forall t:t=t^{\prime }=0)%
\text{ \ \ \ \ \ Center of Rotation\ \ \ \ (b)}  \label{13-centrifugal}
\end{eqnarray}
So the acceleration\textbf{\ }is clearly translated in hyperbolic motion by
an \textbf{elongation} (13a) or extension of the space itself from the fixed
point $O^{\prime }$($\equiv O)$ into $\gamma O^{\prime }P^{\prime }$ in K:
Everything happens as if there was a fundamental \textbf{''elastic''}
elongation $\gamma X$ of distance $X$ . We suggest to call this motion 
\textit{''expanding motion''}. Given that, in non-relativistic rectilinear
accelerated motion, the velocity is intrinsically non-constant, we admit a
fundamental contrast (at one space dimension!) between a relativistic 
\textit{hyperbolic elastic} motion ($\frac{c^{2}}{X})$ and a
non-relativistic \textit{circular} motion ($\frac{v^{2}}{r})$ because we
have in both cases (PCCH) a constant velocity (respectively $v$ and c). In
both cases (\textbf{Fig3}) we have also an acceleration in the direction of
the center $(O^{\prime }\equiv O)$: the \textit{center }of the circle (13b)
with constant distance (Euclidean orthogonality acceleration-velocity, \S
3-2) or the \textit{center} of hyperbola (13b) with constant interval
(Minkowskian orthogonality\ with velocity). In both cases O' does NOT\ move
(in phase 2) with respect to O (unlike EBR, \textbf{Fig2}).

But if the acceleration in Euclidean\ rotation motion and Hyperbolic
Rotation motion have the same direction, they have NOT\ the same sense: the
former is centripetal and the latter is \textbf{centrifugal }(\textbf{Fig3}%
). So the basic expansion is inscribed in LT. Indeed the Principle of
correspondence ''circle-hyperbola'' (PCCH) involves that the physical
hyperbolic motion, induced by $ALT$, must be \textbf{Hyperbolic rotation}
(LT), in geometrical meaning.

Elastic motion is not compatible with Einstein's contraction\ in rigid
motion and this is the reason why we have to give up Einstein's definition
of distance. In order to do that we have to define a passive use of LT that
leads to a dilation of distance (13a). We will prove that the best way to
precise Born's ''boost-invariance'' of the distance is to bring invariance
of \textit{proper distance} into line with invariance of \textit{proper time}
(new symmetry in SR, \S 3-4).

\subsection{Non-zero norm of space-like 4-vector acceleration}

The essential physical data of the material point in pre-relativistic
kinematics are given by the 3-vectors position $\mathbf{\vec{r}}$, velocity $%
\mathbf{\vec{v}}$ and acceleration \textbf{$\vec{a}$}. In relativistic
kinematics, these data are replaced by the space-temporal 4-vectors $\vec{R}$%
, $\vec{U}$ and $\vec{A}$ . The importance of the transition from 3-vector 
\textbf{$\vec{a}$} (pre-relativistic, a Galilean invariant) to that of a
4-vector $\vec{A}$ is completely underestimated. The essential
characteristic of $\vec{A}$ is that it is a spacelike 4-vector, orthogonal,
in the Minkowskian meaning, to the velocity 4-vector $\vec{U}^{\prime }$ ,
which is time like :$\vec{U}^{\prime }\ast \vec{A}^{\prime }=\vec{U}\ast 
\vec{A}=0.$

We thus have the norms $\left\| \vec{U}\right\| =\left\| \vec{U}^{\prime
}\right\| <0$ and\ \ $\left\| \vec{A}\right\| $\ $=$\ $\left\| \vec{A}%
^{\prime }\right\| >0$, that must be \textit{different from 0 (14), } the
zero norm being characteristic of a light-type 4-vector. In the general case
of three space dimensions, the components of $\vec{A}^{\prime }$, $\gamma
^{\prime }(\gamma ^{\prime }\mathbf{a}^{\prime }$ $+$ $\frac{d\gamma
^{\prime }}{dt^{\prime }}\mathbf{v}^{\prime }\mathbf{,}$ $\gamma ^{\prime
3}v^{\prime }.\frac{dv^{\prime }}{dt^{\prime }}c)$ describe the motion of a
particle$\ P^{\prime }$ in K', which is also a motion in K. Let us restrict
the 3-space dimensions problem to that of a single space dimension, in which
case the space-time ''4-vectors'' become space-time 2-vectors $\vec{A}%
^{\prime }(A_{x}^{\prime }$, $A_{t}^{\prime })\overset{LT}{\rightarrow }\vec{%
A}(A_{x}$, $A_{t})$. If the particle $P^{\prime }$ is now a rest (fixed) in $%
K^{\prime }$ ($v^{\prime }=0$, $t^{\prime }=0$) spacelike 4-vector becomes
2-vector acceleration $(a=\frac{dv}{dt},$ $\alpha =a^{\prime }=\frac{%
dv^{\prime }}{dt^{\prime }})$ and invariant norm of acceleration becomes
proper acceleration: 
\begin{equation}
\vec{A}^{\prime }(\alpha ,0)\text{ \ \ \ }\overset{PLT}{\rightarrow }\text{
\ \ \ \ }\vec{A}(\gamma \alpha ,\gamma \beta \alpha )\ \text{ \qquad }\ \
\left\| \vec{A}^{\prime }\right\| =\left\| \vec{A}\right\| =\alpha \text{%
\qquad\ }\alpha _{Boost}\text{ }\neq 0=\gamma ^{3}a  \label{14-non zero norm}
\end{equation}
$(a_{x}=\gamma ^{2}a+\beta a_{t}=\gamma ^{2}a+\gamma \beta ^{2}\alpha
=\gamma \alpha \Longrightarrow \alpha =\gamma ^{3}a).$ In \textbf{Fig3} the
direction of acceleration along $O^{\prime }x^{\prime }$ must be permanently
orthogonal (in Minkowski's meaning) to the direction of velocity. The norm $%
\alpha $ of \textit{spacelike} $\vec{A},$ the proper acceleration in any
Lorentz boost, cannot be equaled to zero (definition of lightlike 4-vector).

\subsection{Proper time on elastic rod and renormalization of Minkowskian
metric}

We now examine the phase of acceleration in order to deduce firstly in K the
duration of acceleration $\Delta t_{P^{\prime }}$ and, from the elongated
distance, the distance of acceleration $\Delta x_{P^{\prime }}$

\begin{equation}
\Delta t_{P^{\prime }}=\beta \gamma X_{P^{\prime }}\qquad \qquad \ \ \
\Delta x_{P^{\prime }}=\gamma X-X=X(\gamma -1)
\label{15-acceleration duration}
\end{equation}
with kinematics form $\alpha \Delta x_{P^{\prime }}=\gamma -1=\frac{v^{2}}{2}%
+...$ (Taylor) which corresponds to \textit{relativistic kinetic energy }$%
mc^{2}(\gamma -1)$ of the particle $P^{\prime }$. We obtain a different
duration of acceleration for any point $X$ of the rod\ ($0\leq t\leq \beta
\gamma X)$ $\ $We now calculate, by a simple integration, this duration 
\textit{in proper time} of the Lorentz boost because we have $t^{\prime }=0$
but $dt^{\prime }\neq 0$ with $dx^{\prime }=0$ and with (8) (in
Born-Rindler's notation we have in our theory for $Ox$ hyperbola: $dX=0$,
note 3)

\begin{equation}
\tau _{P^{\prime }}=\int dt^{\prime }=\int d\tau =\int_{0}^{t_{P^{\prime }}}%
\sqrt{1-\beta ^{2}}dt=\int_{0}^{t_{P^{\prime }}}\sqrt{1-\frac{\alpha
^{2}t^{2}}{1+\alpha ^{2}t^{2}}}dt=\frac{1}{\alpha }\func{arcsinh}\alpha
t_{P^{\prime }}  \label{16-proper time}
\end{equation}
The transition between $0$ and $\beta $ velocity will be shorter $\tau
_{P^{\prime }}$ ($\approx \frac{1}{\alpha }\ln 2\alpha t_{P^{\prime }})$ in
proper time in $K^{\prime }$ than in time of $K$: \ $\tau _{P^{\prime
}}<t_{P^{\prime }}.$ By using \textit{hyperbolic velocity} 
\begin{equation}
\beta _{H}=\ln \sqrt{\frac{1+\beta }{1-\beta }}=\alpha \tau _{P^{\prime }}
\label{17-hyperbolic velocity}
\end{equation}
(the well known parameter of rapidity) we rediscover the hyperbolic
coordinates of $Ox$ hyperbola $(t^{\prime }=0)$ 
\begin{equation}
t_{P^{\prime }}=\frac{1}{\alpha }\sinh \alpha \tau _{P^{\prime }}=\frac{1}{%
\alpha }\sinh \beta _{H}=\beta \gamma X\qquad \ \ \ \ x_{P^{\prime }}=\frac{1%
}{\alpha }\cos \alpha \tau _{P^{\prime }}=\frac{1}{\alpha }\cos \beta
_{H}=\gamma X  \label{18-hyper-coord}
\end{equation}
The proper time $\tau _{P^{\prime }}$ in order to reach the cruise velocity
is a different one in each point of the rod in the phase of acceleration
(phase 2). We have therefore the joint between the second phase and the
third phase, the second limit of integration (because $t^{\prime }=0$ but $%
dt^{\prime }\neq 0$):

\begin{equation}
t^{\prime }=0\text{ but }dt^{\prime }\neq 0\qquad \Longrightarrow \qquad
t^{\prime }=t-\beta x=\tau -\beta _{H}X=0  \label{19-basic time equation}
\end{equation}
We rediscover the instantaneity $\tau _{O^{\prime }}=0$ for one end $%
O^{\prime }$ of the rod and the maximal proper time $\tau _{P^{\prime }}$
for the other end of the rod. We have the following coupling for any
Hyperbolic Rotation (HR) or any LT 
\begin{equation}
\text{ }d\beta _{H}=d\tau =\gamma ^{2}d\beta \text{ \ \ \ (a) \
(differential hyperbolic angle)}\qquad \text{\ }\beta _{H}=\alpha \tau \text{
\ \ \ \ (b) \ (integral hyperbolic angle)}  \label{20-glocal}
\end{equation}
given that $\beta _{H}=\int d\beta _{H}=\int \alpha d\tau =\alpha
\int_{0}^{t_{P^{\prime }}}\sqrt{1-\beta ^{2}}dt=\alpha
\int_{0}^{t_{P^{\prime }}}\sqrt{1-\frac{\alpha ^{2}t^{2}}{1+\alpha ^{2}t^{2}}%
}dt=\func{arcsinh}\alpha t_{P^{\prime }}=\alpha \tau _{P^{\prime }}$. Unlike
Rindler (11, note 2) we admit that the proper time, defined $\tau $ by (19)
is the fundamental parameter on the rod. \ Let us remark that in proper time
basic equation is of first order whose solution is an exponential (33) at
the condition that $\alpha \neq 0.$ So with our parameter $\tau $ we obtain
naturally a Minkowskian metric $\ (1,$ $\ -1)$ in phase2. 
\begin{eqnarray}
dt^{2}-dx^{2}\text{ } &=&d\tau ^{2}-(dx^{\prime }=0)^{2}=\alpha ^{-2}d\beta
_{H}^{2}\text{\qquad }\rightleftarrows \qquad x^{2}-t^{2}=\alpha ^{-2}\text{%
\ \ \ \ signature\ }(1,\ -1)  \label{21-undetermined} \\
\alpha &=&0\text{ \ }\Longrightarrow \text{ \ }d\beta _{H}=0\text{ \ \ \ \ }%
\Longrightarrow \text{ \ \ \ }dt^{2}-dx^{2}\text{ }=\infty .0=\text{%
underterminate \ }  \notag \\
\text{\ }x^{2}-t^{2} &=&\infty =x^{2}\text{ \ (non relativistic)}  \notag
\end{eqnarray}
However we know that there is an hidden boost $\alpha $ behind the ''arc of
hyperbola'' in Minkowski's metric with $ALT$ (\textbf{Fig3}). This
unexpected result is geometrically obvious because if you take a point at
rest and you apply a series of $\alpha LT$ \ between two limits of
integration (basic equation 19) the successive points will be placed not a
straight line but... on an arc of hyperbola. Let us note that in SSR the
infinitesimal interval is undeterminate if $\alpha =0$ (Minkowski's standard
metric is a non-calibrated metric and need a course of treatment (34) of
renormalization (note 1) !

\subsection{Synthesis between PLT and ALT and coupling ''Born's acceleration
and Minkowski's distance''}

We come back now to Minkowski's hyperbolas (beginning of this paper). We
focus the attention on the demi-branch (\textbf{Fig1}) of hyperbola $%
x^{2}-t^{2}=x^{\prime 2}=d^{2}\ (x>0,$ $t\geq 0).$\ The distance $d$ is
defined by the interval between two simultaneous events in $K^{\prime
}(t^{\prime }=0)$

\begin{equation}
\vec{R}^{\prime }(d,0)\overset{PLT}{\rightarrow }\vec{R}(\gamma d,\text{ }%
\gamma \beta d)\qquad \ \ \left\| \vec{R}^{\prime }\right\| =\left\| \vec{R}%
\right\| =d\qquad \text{\ \ \ \ \ \ \ \ }x=\gamma d
\label{22-Minkowski's distance}
\end{equation}
Obviously $d$ in K is no longer a ''distance'', in the usual meaning, but an
interval. However, if we consider that a difference of space coordinates $x$
in K defines a distance, then such a distance $x=$ $\gamma d$ is defined
between two \textit{non-simultaneous events} in K (gap of simultaneity: $%
\gamma \beta d)$. This is very interesting for astrophysical and
cosmological distances \S 3-5). It is clearly a non-Einsteinian definition
of (the dilation of) distance. However this new definition is induced from a
perfect \textit{symmetry in }$PLT$\textit{\ or }in Minkowski's hyperbolas
when we replace ''events at the same place'' by ''events at the same time''
in K' and ''duration'' by ''distance'' in $K-K^{\prime }$. The duration $D$
between ''two events not at the same place'' or the distance $d$ between
''two events not at the same time'' is dilated in K (\cite{11b}). Both $%
\alpha LT$ and $dLT$ induce a dilation of distance. Minkowski's and Born's
arc of hyperbola (\textit{if }$d=X$\textit{)}, \textit{\ are completely
identical and so the couple (}$\alpha _{Born}$, $d_{Minkowski})$\textit{\ } 
\begin{equation}
X_{\text{Born}}=d_{\text{Minkowski}}\qquad \Rightarrow \qquad d=\alpha
^{-1}\qquad \qquad \alpha LT\equiv dLT  \label{23-coupling one}
\end{equation}
has to be a fundamental relativistic couple.\textit{\ }Unlike
prerelativistic kinematics, in which there is no fundamental relationship
between $\mathbf{\vec{a}}$ and $\mathbf{\vec{r}}$ $,$ there is a
relativistic four-vectorial relationship () 
\begin{equation}
(d\text{, }0)(\alpha ,0)=\alpha _{\text{Born}}d_{\text{Minkowski}}=1\text{ }%
\qquad \qquad \vec{A}^{\prime }.\mathbf{\ast }\vec{R}^{\prime }=\mathbf{\vec{%
U}}^{2}=c^{2}=1\text{ \ \ (b)}  \label{24-coupiling two}
\end{equation}
We verify with $\alpha t=\beta \gamma $ and \ $\alpha x=\gamma :\ \vec{A}%
\mathbf{\ast }\vec{R}=(x$, $t)(\gamma \alpha $, $\gamma \beta \alpha
)=\alpha ^{2}x^{2}-\alpha ^{2}t^{2}=\alpha ^{2}x^{2}-\alpha ^{2}t^{2}=1.$
The factor onehalf $\frac{1}{2}$ ($ad=\frac{1}{2}v^{2}$ $\Longrightarrow E=%
\frac{1}{2}mv^{2})$ disappears in a relativistic theory ($ad=c^{2}\
\Rightarrow \ E=mc^{2})$.

\subsection{Emission of radiation, Poincar\'{e}'s definition of distance $d$
and Doppler redshift}

According to Rindler ''the rod ends in a photon''. But in Born-Rindler's
theory a photon dispatched to chase the particle at $t=0$ from $O$ never
catch up with it because the acceleration remains constant. In our new
theory of Einstein's boost, the acceleration stops in the third phase and
there is an intersection $M^{\prime }$ between the tangent and the wordline
of the photon. In CR, Einstein's boost is structurally an emitter of
radiation and this emission is not independent of the definition of distance.

\subsubsection{Synchronized clocks and length L (Einstein 1905): rigid
system and rigid rod}

In the $K^{\prime }$inertial system, let us first consider Einstein's rigid
rod $O^{\prime }M^{\prime }$ at rest in K(of length L) with a source at $%
O^{\prime }$ and a mirror at $M^{\prime }$. A light signal is emitted from $%
O^{\prime }$ at $t^{\prime }=0$, it is reflected in $t^{\prime }=\tau /2$ at 
$M^{\prime }$ and returns to $O^{\prime }$ at $t^{\prime }=\tau =2L/c$, the
''time out''$\frac{\tau }{2}$ being equal to the ''back time'' $\frac{\tau }{%
2}$. Einstein's clock synchronization uses three successive physical events,$%
1,2,3$: $O^{\prime }(0,0)_{1},$ $M^{\prime }(L,\frac{\tau }{2})_{2}$,\ $%
O^{\prime }(0,\tau )_{3}$. He has synchronized the two clocks at the ends of
the rod and defined the simultaneity of two events ''at a distance''. These
two simultaneous events $(0,$ \ $\frac{\tau }{2})_{2}$ \ and \ $(L,$ $\frac{%
\tau }{2})_{2}$, respectively at $O^{\prime }(x_{O^{\prime }}^{\prime }=0)$
and $M^{\prime }(x_{M^{\prime }}^{\prime }=L)$ are however not explicitly
given in Einstein's paper ($\tau $ represent here Minkowski's standard
proper time).

\subsubsection{Synchronous distance d (Poincar\'{e} 1908): abstract systems
and light-distance}

The same system is used, but without Einstein's rigid rod (given a priori)
and with a single clock in $O^{\prime }$. We will suppose that the mirror $%
M^{\prime }$ is at rest in $K^{\prime }$ (a distant reflecting object) at an
unknown distance. A light signal is emitted at $O^{\prime }$ at $t^{\prime
}=0$, reflected in $M^{\prime }$ in $D$ (Duration) and returns to $O^{\prime
}$ at $t^{\prime }=2D$. The proper, ''synchronous'' distance $d$ is defined
by two simultaneous events in K':

\begin{equation}
\text{\ simultaneity: \ :}(0,\text{ \ }D)_{2}\text{ and\ }(d,\text{ }D)_{2}%
\text{ \ \ \ \ }\ (d=D)_{Poincar\acute{e}}=\frac{1}{2}\tau _{Minkowski}\text{
(round trip)}  \label{25-Bondiround trip}
\end{equation}
The ''synchronous'' distance $d$ may be measured by a single clock in $%
O^{\prime }$ ($2D$) with a round trip signal. Such a distance, measured by
the light time of travel , is not a new technique. We can replace Einstein's
rigid rod by ''one half light travel time (\textit{two ways or round trip }$%
2D$) distance'' with \textit{Bondi's radar method} (\cite{13}). We add a new
element: the converse (Poincar\'{e}) interpretation of Einstein's
synchronization involves also that the so formed distance $d$ is transformed
by LT $\gamma d$ like the duration $\gamma D$. In other words, $d$ is a
distance that is defined as the proper duration, and is thus the \textbf{%
shortest} in K'. As a last analysis and from a historical viewpoint,
Einstein's work is based on the direct theorem (the $O^{\prime }t^{\prime }$
axis), while Poincar\'{e}'s opened the way to the reciprocal (the $O^{\prime
}x^{\prime }$ axis or $t^{\prime }=0$). \textit{''This Lorentz hypothesis is
the immediate translation of Michelson's experiment, if the lengths are
defined by the time that light takes to travel through them''} (\cite{10} , 
\cite{11} \& \cite{11b}). In the same way that ''simultaneity at a
distance'' cannot be defined independently from the velocity of light, the
distance itself may not be defined independently from this velocity. This 
\textit{proper} distance $d$ (events at the same time $t^{\prime }=0$ in K')
then becomes an ''\textit{invariant''} of LT ($x^{2}-t^{2}=d^{2}=s^{2})$ in
the same sense as the proper duration $D$ (events at the same place $x=0$ in
K') is an ''invariant'' of LT ($t^{2}-x^{2}=\tau ^{2}=s^{2})$. \textit{In
summary Poincar\'{e}'s proper length (hyperbola along }$\mathit{Ox}$\textit{%
) is the exact symmetric of Minkowski's proper time (hyperbola along (Ot);
except the one-half }$\frac{1}{2}$\textit{\ factor (two-ways) in (25). }

\subsubsection{One way light travel distance and Bondi's Doppler redshift
factor (phase 3)}

We remark now that the proportionality duration-distance, according to
Poincar\'{e}'s basic elongated ellipse\cite{10}, is also valid for ''one
way'' trip (without factor $\frac{1}{2})$. We have in K because when the
phase of acceleration stops (phase 3), O' is in motion with respect to O: 
\begin{equation}
(\sqrt{\frac{1-\beta }{1+\beta }})_{forth}+(\sqrt{\frac{1+\beta }{1-\beta }}%
)_{back}=2\gamma \text{ \ \ (round trip) \ \ \ \ \ \ \ }d_{bzck}=kd(=D)\text{
\ }  \label{26-round trip}
\end{equation}

So the factor defining the distance by the travel time back is no longer $%
\gamma $ $\in \lbrack 1,\infty $ $[$ but famous Bondi's factor $k$ $=\sqrt{%
\frac{1+\beta }{1-\beta }}=e^{\beta _{H}}\in $ $[1,\infty \lbrack .$ Both
factors are Lorentz factor $\beta =\tanh \ln k=\frac{k^{2}-1}{k^{2}+1}$ \ ($%
\frac{t}{x}=\beta )$. However in Bondi's theory, this factor is introduced
in the transformation of the lengthwave and not the transformation of length
itself: it is a Doppler factor. Given that, any one-way length is determined
by the one-way light travel time, the transformation of \textit{length }and 
\textit{wavelength} are the same. In order to define the distance we must
have a source in P'.\ Suppose a monochromatic source of lengthwave $\lambda
^{\prime }$ in $P^{\prime }(K)$. Observed at O in $K$ we have the lengthwave 
$\lambda $

\begin{equation}
\frac{d_{bzck}\text{{\small (reception)}}}{d\text{({\small emmission)}}}=%
\sqrt{\frac{1+\beta }{1-\beta }}=\text{\ }\frac{\lambda \text{{\small %
(reception)}}}{\lambda ^{\prime }\text{{\small (emmission)}}}=k\ 
\label{27-Doppler}
\end{equation}
We rediscover with Poincar\'{e}'s dilation of distance the same Doppler
formula as Einstein's one (longitudinally, the direct calculus with LT is
very easy, \cite{11b}). But with the theory of Lorentz boost we have
structurally a redshift ($\beta _{H}=\ln k$ and not $\beta _{H}=\ln k^{-1})$%
. We suppose here that the acceleration stops (phase 3) and so we have (the
derivation with respect of the proper time $\tau $ is primed) 
\begin{equation}
\alpha =\beta _{H}^{\prime }=\frac{k^{\prime }}{k}=0\text{ \ \ \ \ \ \ \ }%
\Rightarrow k(\tau )=cte  \label{28-Bondi}
\end{equation}
So in the framework of standard SR Bondi's factor is not an expansion factor
(a scale factor) because it does not depend of proper time (see CR).
Finally, we remark also that the one-way outgoing signal is still not used
until now.

\section{Fundamental Lorentz boost, minimal acceleration and one-way maximal
distance in CR}

\bigskip Now we are able to transform SR into CR (\S 1). Until now we have a
fundamental relationship between acceleration in Lorentz boost ($\alpha LT$)
and distance ($dLT$) defined by interval in Minkowski's space-time (23). The
cause of this acceleration is not inscribed in the structure of space-time
and when the cause (the force) no longer acts (M'), the acceleration stops $%
\alpha =0=\Lambda $ and we rediscover the standard SR with uniform cruise
velocity or the straight line geodesic. In vacuum flat space-time we cannot
place ''on the same plane'' $(x,$ $t)$ the $ALT$ and the $PLT$. Without a
fundamental acceleration $\alpha _{\min }$ there is no Lorentz boost, no
active LT, no hyperbola, and finally no complete SR. Without hyperbola, it
remains the light cone, with only one invariant in standard SR. If we wish
take into account the hyperbola, we have to consider the existence of a
minimal (\cite{3a}) acceleration in the ''vacuum''. 
\begin{equation}
a_{\min }\neq 0=a_{boost}\Longrightarrow \qquad a_{\min }d_{\text{horizon}%
}=a_{M}R_{H}=c^{2}=1\text{ \ \ \ \ }\qquad \Longleftrightarrow \qquad
x^{2}-t^{2}\text{ \ \ }{\LARGE \nrightarrow }\text{ \ }\infty
\label{29-Horizon}
\end{equation}

The existence of $a_{M}$ (24) involves the existence of a fundamental
Lorentz boost structurally connected with an emission of radiation. Given
that distance $d$ is measured in lightyear, if we send a light signal
towards a galaxy we cannot wait for the return of the light. Our new
definition of \textit{proper distance}\ $d$ \ supposes the existence of a ''%
\textit{one way'' (without one half) }$t^{\prime }=D_{\text{one-way}}$ light
travel time'' distance, i.e. a \textit{non-infinite including unit} in
hyperbola (2). The radius of Hubble $R_{H}$ is precisely defined by the
one-way travel lifetime $T_{\text{one-way}}$ of the universe or time of
Hubble $T_{H}$ (emission $CBR$ cosmological background radiation, or photons
structurally boosted fifteen billions years ago(\textit{fundamental
acceleration involves fundamental emission of CBR}). With Poincar\'{e}'s
distance (25) we have: ($t=0$, emission) \ \ 
\begin{equation}
\text{\ simultaneity: }(0,\text{ \ }T_{H})_{2}\text{ \ and\ \ \ }(R_{H},%
\text{ }T_{H})_{2}\qquad \ \ \ \ \ R_{H}=T_{H}\text{ (one way) \ }
\label{30-one way}
\end{equation}
Bondi's mirror $M^{\prime }$ becomes Penzias-Wilson antenna,\textbf{\ Fig5}).

. So with the basic including hyperbolic unit we have now

\begin{equation}
d_{R_{H}}=r<R_{H}\text{ \ \ \ or \ \ \ \ \ }r=\varepsilon R_{H}\qquad 0\leq
\varepsilon \leq 1  \label{31-epsilon}
\end{equation}

We will show that $\varepsilon =\beta .$ \ On the other side, the
introduction of a minimal acceleration $a_{M}$ involves the deletion of the
third phase of cruise velocity: any body in the vacuum undergoes a basic $%
a_{M}-HR$. The sound Pioneer $P^{\prime }$ follows in curved space-time a 
\textbf{geodesic} which is a generalization of inertial principle

\subsection{Milgrom's acceleration and cosmological acceleration of expansion%
}

What is the relationship between Milgrom acceleration $a_{M}=H$ $(c=1)$ and
the parameter of acceleration $q_{0}$ in Cosmology? In other words what is
the relationship with our constant Hubble radius $R_{H}$ and the ''ad hoc''
scale factor in Friedman's metric (Robertson-Walker's metric), generally
noted $a(t)$ (with derivatives with respect to time, $\dot{a}(t)$, $\ \ddot{a%
}(t)$ and with $q_{0}=-\frac{a;\dot{a}}{\ddot{a}^{2}})$? In CR this factor
is naturally equaled to $q_{0}=-1$ because we have $\frac{a_{M}R_{H}}{c^{2}}%
=1.$ So we have then an accelerated expansion in SR. For the remote galaxies 
$r\lessapprox R_{H}$ we obtain the same result with the law of expansion in $%
k$ (factor of Bondi) ($1+z$ )$_{\text{Friedman}}$ $=1+\beta +(1+\frac{q_{0}}{%
2})\beta ^{2}+...=k_{\text{CR}}=\sqrt{\frac{1+\beta }{1-\beta }}=1+\beta +%
\frac{1}{2}\beta ^{2}+....$ So we have:

\begin{equation}
a_{M}=H(c=1)\qquad \Longrightarrow \qquad q_{0}=-1  \label{32-acceleration}
\end{equation}

Unlike in Robertson-Walker's metric, we does not need an ad hoc factor in CR
because Hubble constant \ $H$ is now explicitly connected with the
logarithmic derivation of expanding Bondi factor $k$ in proper time
(primed). Indeed fundamental equation is in CR is $H=\frac{d\beta _{H}}{%
d\tau }$ and so the contrast 
\begin{eqnarray}
H &=&\frac{\dot{a}(t)}{a(t)}\text{ (Friedman-Lema\^{i}tre)}\qquad \ \ \ \ \
\ \ H=\beta _{H}^{\prime }=\frac{k^{\prime }(\tau )}{k(\tau )}=\alpha _{M}%
\text{ \ (CR)\ \ \ \ }  \label{33-Bondi two} \\
&\Rightarrow &\text{ \ }k(\tau )=k_{0}e^{H\tau }=e^{H\tau }  \notag
\end{eqnarray}
because $k(0)=1$ in the boost. And so the non nul Hubble constant involves
an exponential expansion compatible with hyperbolic constant horizon. In
summary we have switched to the second member $dt^{2}-dx^{2}=a(t)=Cte$
Friedman's scale factor ($dt^{2}-a(t)dx^{2}$) by transforming it into
constant. Then we showed that the scale factor of expansion is Bondi's
factor $k$ and we have therefore to discover a relationship between $R_{H}$
and $k$ (36).

\subsection{\protect\bigskip Renormalized Minkowski's metric, hyperbolic
horizon and global Hubble's law}

The Lorentz boost describes the continuous transformation from any point of
the elastic rod from the velocity $0$ to $\beta .$\textit{\ }The element of
proper time $d\tau $ depends in CR on Hubble constant: With $\alpha _{Min}$
we have a renormalized Minkowski's metric (21) by introducing a fundamental
unit $R_{H}$ \ 
\begin{equation}
\text{\ \ \ }dt^{2}-dx^{2}\text{ }=d\tau ^{2}=H^{-2}d\beta
_{H}^{2}=R_{H}^{2}d\beta _{H}^{2}\text{ \ \ \ \ \ \ }\frac{d\beta _{H}}{%
d\tau }=\frac{k^{\prime }\text{ }}{k}\text{\ \ \ \ \ \ (LOCAL)}
\label{34-normalmetric}
\end{equation}
\ 

Locally for\textit{\ }$r<<R_{H}$\textit{\ (}$\alpha >>\alpha _{Min})$ $%
\allowbreak $standard flat metric of SSR is still valid. Indeed locally the
hyperbola becomes a parabola $x=\frac{c^{2}}{a}(1+\frac{\alpha ^{2}t^{2}}{%
2c^{2}}+...)=\frac{c^{2}}{a}+\frac{1}{2}\alpha _{_{0}}t^{2}+$ ) and
parabolic Einstein's principle of equivalence on the basis of GR is still
valid but \textit{globally} we have the curvature of fundamental Milgrom's
(demi-branch of) hyperbola $\alpha =\varrho $ (\textbf{Fig5}).

\begin{equation}
x^{2}-t^{2}=R_{H}^{2}\text{ \ \ \ \ \ \ \ \ \ }\beta _{H}=H\tau =\ln k\text{%
\ \ \ \ \ \ (GLOBAL) }  \label{35-global}
\end{equation}

Let us finally show that the geometry is hyperbolic in the meaning of Cayley
and Klein.. With Cayley-Klein's hyperbolic distance $r_{H}$ induced from the 
\textit{cross-ratio} formula is given by $r_{H}=r_{h}\func{arctanh}\frac{r}{%
r_{h}}=\ln \sqrt{\frac{1+\frac{r}{r_{h}}}{1-\frac{r}{r_{h}}}}$where $r$, is
a Cartesian distance smaller than $r_{h}$ the horizon. We have from the
global form $\tau _{P^{\prime }}=\frac{1}{H}\ \beta _{H}=T_{H}\ln \sqrt{%
\frac{1+\beta }{1-\beta }}$ \ and therefore $\tau _{P^{\prime }}=T_{H}\ln 
\sqrt{\frac{1+\frac{\beta T_{H}}{T_{H}}}{1-\frac{\beta T_{H}}{t_{h}}}}.$
With $\xi _{P^{\prime }}=c\ \tau _{P^{\prime }}(c=1)$ and $R_{H}=cT_{H}$ $%
(c=1)$\ ($\xi _{P^{\prime }}$ is the proper distance\footnote{%
The element of proper distance $d\xi $ is connected by the other hyperbola
Ot (Fig1). We have a perfect symmetry between proper time $d\tau =dt^{\prime
}$ and proper length $d\xi =dx^{\prime }$ with the other ($Ot)$ hyperbola $%
t^{2}-x^{2}=T_{H}^{2}$ and $d\xi ^{2}=dx^{2}-dt^{2}$ if $d\xi =id\tau .$}
with element of proper distance $d\xi _{P^{\prime }}=dx^{\prime }$), we have
a basic relationship between $k$ and $R_{H}$ 
\begin{equation}
\tau =T_{H}\ln \sqrt{\frac{1+\frac{\beta T_{H}}{T_{H}}}{1-\frac{\beta T_{H}}{%
t_{h}}}}=\ln k\qquad \Rightarrow \qquad \xi =R_{H}\ln \sqrt{\frac{1+\frac{%
\beta R_{H}}{R_{H}}}{1-\frac{\beta R_{H}}{R_{H}}}}=R_{H}\ln \sqrt{\frac{1+%
\frac{r}{R_{H}}}{1-\frac{r}{R_{H}}}}  \label{36-Klein}
\end{equation}
where $k$ is the redshift Doppler factor of Bondi \cite{13}. This is the
inverse form of (33). And so we deduced, with $r\lessapprox R_{H}$ the law
of Hubble ($0$ $\leq \beta \leq 1)$

\begin{equation}
\beta =Hr  \label{37-Hubble}
\end{equation}
From integral form , we can\ therefore deduce globally the law of Hubble. In
CR, like in Esher's drawing, there is a constant hyperbolic horizon and an
expanding Universe with potentially infinite hyperbolic distances, $%
(x\rightarrow \infty $ or $\xi \rightarrow \infty $) in time $t$ and proper
time $\tau $ as well. The Universe in CR (33-36) is not static but in
''steady state''\cite{13}. Nothing is changed for a light point because we
have obviously $ds^{2}=d\tau ^{2}=dt^{2}-dx^{2}=0.$ Is the light sensitive
to the curvature in CR or to cosmological constant $\Lambda $ (\cite{2d})
Given that light follows in ''vacuum'' a straight line and that straight
line is in CR hyperbolic straight line (with curvature), the answer is
therefore positive.

\section{Conclusion:\ Einstein's cosmological constant and Poincar\'{e}'s
negative pressure of the ''vacuum''}

\bigskip Our main heuristic principle is the principle of correspondence
non-relativistic and relativistic kinematics (PCCH). We have the theater of
circular motion and the theater of hyperbolic rotation motion. The roles but
not the actors are the same. Let us give an example. Who plays the role of
angular velocity $v=R\omega $? In uniform circular motion, unlike constant
angular velocity $\omega $ (in standard units $s^{-1}$), linear velocity $v$
depends on radius. In hyperbolic motion linear velocity depend on the radius 
$c=R_{H}H$ unlike Hubble constant (in standard units $s^{-1}$) which does
not depend of distance (37) $\beta =rH.$ \ 

Last but not least: Who plays the role of the transcendent $\pi $ in\
hyperbolic CR? The fundamental differential equation in CR, based on $%
a_{M}-HR$, is a first order equation (33-36). Thanks to $\Lambda \neq 0$ the
solution is not trivial and CR is exactly the completely hyperbolic theory
of hyperbolic rotation, i.e. for Lorentz boost with Milgrom's acceleration.
The circular motion governed by $\pi $ is replaced in CR by an expanding
hyperbolic motion governing by $e$. This complete substitution characterizes
the aesthetic superiority of hyperbolic geometry on the other non-Euclidean
geometries \cite{14}.

Einstein had introduced 1n 1917 the cosmological constant $\Lambda $ in the
first member of Einstein's GR equation; if it is switched to the second
member, as is now the general practice, it becomes a term independent from
matter in the usual sense of the word. But what is the meaning of a
constant, unrelated with matter, in a theory (GR) which by essence gives a
structure based on matter to the space-time? We suggest, as a consequence,
not to shift it from one member to the other, but to change its relativity
in order to reinsert it into its natural ''environment'', in other words, in
the ''vacuum'' space-time of Minkowski.

Nothing is changed with GR except that $\Lambda $ must be introduced in
Einstein's equation. The cosmological constant, $\Lambda $, when it is
written on the right side, introduces the constant global curvature of the
universe (\cite{6}), as distinct from all the other terms of Einstein's GR
equation, which correspond to the local curvature (defined by gravitation in
each point of the space-time). The fact that the acceleration may not be
null is not in contradiction with the local Riemann flatness of the
pseudo-Euclidean space, as derived from the affine connection (the
''Christoffel''), under the condition that $\Lambda $, defined by CR, be
introduced into Einstein's GR equation.

Our use of inverted commas in the title of our paper means that the
classical ''vacuum'' is not really empty because there is a fundamental
field that can be called ''Poincar\'{e}'s relativistic aether'' or ''hidden
relativistic fluid'' or in ''dark energy'' (it's the fashion). Given that
nothing is changed between the relation with GR except that the cosmological
constant is imposed by CR, we can determine thanks to Einstein's GR, and
more precisely from Einstein's static model of Universe, the very weak
density of Poincar\'{e}'s fluid with Poincar\'{e}'s \textit{%
non-electromagnetic} negative pressure (\cite{10})

\begin{equation}
\rho _{\text{hidden fluid}}=\frac{\Lambda _{\text{Einstein}}}{4\pi G}=-p_{%
\text{Poincar\'{e}}}  \label{38-Poincareinstein}
\end{equation}

If Einstein is the author of the very well known $E=mc^{2}$, Poincar\'{e} is
the author of the very unknown $p=-\rho c^{2}$ in the framework of LT.
Einstein considered at the end of his life that the cosmological constant $%
\Lambda $ in 1917 was his ''greatest mistake''. Maybe Poincar\'{e} thought
also, at the end of his life, that his ''greatest mistake'' was keeping
aether in his relativistic dynamics of 1906 (a relativistic aether but an
aether)? According to our last formula, maybe both, the ''old Poincar\'{e}''
and the ''old Einstein'', were wrong. If we put Einstein's cosmological
constant in his own relativistic kinematics without aether, we rediscover
Poincar\'{e}'s relativistic aether (La ''structure fine'' de la
Relativit\'{e} Restreinte, \cite{12}). This is the irony of the history.

\begin{acknowledgement}
I thank ''mon petit chou'' Pascal Serrano
\end{acknowledgement}



\section*{Supplementary material}

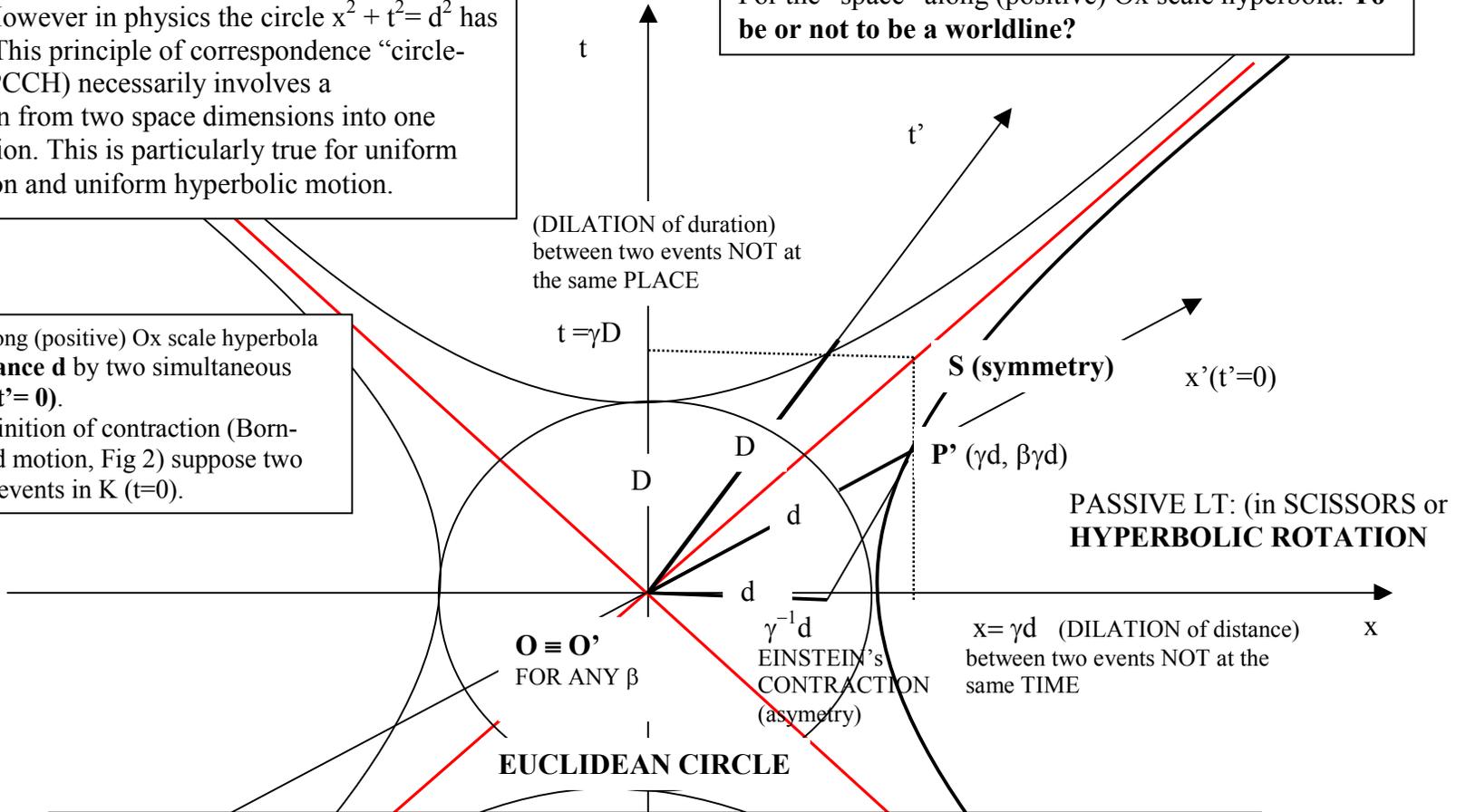

**FIG 1**
**Minkowski's hyperbolas** (scale or calibration hyperbolas, c=1)
Invariance of *finite* Interval or finite unit
Invariant proper duration D, $t^2 - x^2 = D^2$ and invariant proper distance d: $x^2 - t^2 = d^2$

The essence of Minkowski's space-time is the transformation of prerelativistic circle into relativistic hyperbolas. However in physics the circle $x^2 + t^2 = d^2$ has no meaning. This principle of correspondence "circle-hyperbola" (PCCH) necessarily involves a transformation from two space dimensions into one space dimension. This is particularly true for uniform circular motion and uniform hyperbolic motion.

For the "space" along (positive) Ox scale hyperbola: **To be or not to be a worldline?**

(DILATION of duration) between two events NOT at the same PLACE

$t = \gamma D$

The "space" along (positive) Ox scale hyperbola defines a **distance d** by two simultaneous events in K' **(t'= 0)**.
Einstein's definition of contraction (Born-Rindler's rigid motion, Fig 2) suppose two simultaneous events in K (t=0).

**S (symmetry)**   x'(t'=0)

**P'** ($\gamma d, \beta \gamma d$)

PASSIVE LT: (in SCISSORS or **HYPERBOLIC ROTATION**

$x = \gamma d$  (DILATION of distance) between two events NOT at the same TIME

**O ≡ O'**
FOR ANY β

$\gamma^{-1} d$
EINSTEIN's CONTRACTION (asymetry)

**EUCLIDEAN CIRCLE**

Minkowski's flat metric in COSMOLOGY is characterized by the complete disappearance (d→∞) of the along Ox hyperbola. Only the asymptotes survive to the extinction of hyperbola. So standard Minkowski's "metric" is a "non-calibrated metric in "non-hyperbolic" standard SR (SSR, only one invariant). Minkowski's calibration hyperbolas: "d→ ∞" does not generate a relativistic invariant. The infinite interval is the Absolute Space in which geodesics are reckoned to be straight worldlines. The "vacuum" is empty. Global Minkoswski's scale hyperbola $x^2 - t^2 = d^2$ (t'=0) has nothing to do, at the first sight, with Minkowski's element of metric $dt^2 - dx^2 = d\tau^2$ (t'=0)

# FIG 2
# Einstein-Born-Rindler's rigid motion (EBR)
## (hyperbolic uniformly accelerated motion of particle P' or rigid rod O'P'=X)
### ("Boost invariance" of rigid rod
### Relativistic model of the non-relativistic RECTILINEAR ACCELERATED MOTION)

Hyperbolic worldline in parametric coordinates
$$x = \gamma X \quad t = \beta\gamma X$$
and Einstein's asymetry
$$x_{Einstein} = \gamma^{-1} X$$
Hyperbolic worldline in Cartesian coordinates
$$x^2 - t^2 = \alpha^{-2} = X^2$$
and Einstein's quadratic form underlying contraction
$$x^2 = x'^2 - t'^2$$
Hyperbolic worldline in Hyperbolic coordinates
$$x = X ch\beta_H \quad t = X sh\beta_H$$
**and RINDLER**'s parameter of time T constant on rigid rod
$$T_{RINDLER} = \beta_H$$

**RINDLER**'s representation of the contraction of rod $\gamma^{-1}X$

The rod "ends in photon" (X>0, x'>0)

$\alpha^{-1}$

O'$_{RINDLER}$ ($\beta$)      P' ($\gamma X, \beta\gamma X$)

ACTIVE LT FOR ANY POINT of the rod

**O ≡ O'**
FOR β=0

$X = \alpha^{-1}$    P' (X, 0)

t', x' (t'=0)

Rindler's boost invariance of "proper length" is not "on the same plane" as Lorentz invariance of proper time. PLT (uniform motion) and ALT (with acceleration) are not "on the same plane". The former is basic whilst the latter is an application for particular case of constant acceleration. **Let us suppose now that the acceleration on O' stops at the time t= βγ X (Fig 3).**

Unlike Minkowski's scale hyperbola in standard SR, Born's hyperpola is deduced integration from a differential equation $\alpha dt = d(\beta\gamma)$ by a double integration. We can also deduce $\alpha\tau = \beta_H$ from $\alpha d\tau = d\beta_H$ or from (dt'=d$\tau$) $d\tau^2 = dt^2 - dx^2$ by a simple integration.

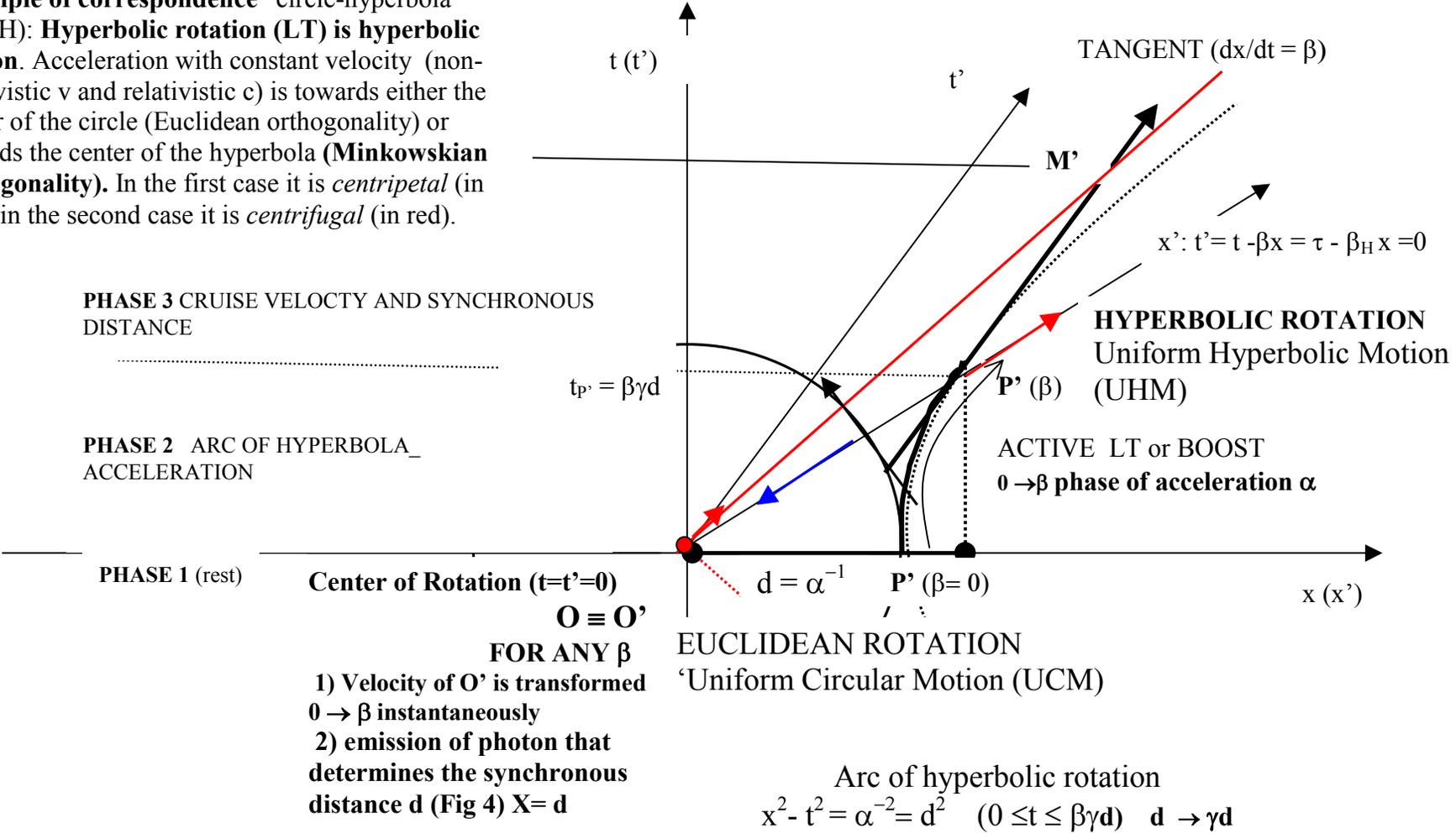

**FIG 3**
**Theory of Einstein's boost (acceleration) with emission of light signal**
"ARC of hyperbola" or "transition of velocity from 0 to β"
**Hyperbolic rotation (geometry, LT) ↔ Hyperbolic motion (physics, BOOST)**

**Principle of correspondence** "circle-hyperbola" (PCCH): **Hyperbolic rotation (LT) is hyperbolic motion**. Acceleration with constant velocity (non-relativistic v and relativistic c) is towards either the center of the circle (Euclidean orthogonality) or towards the center of the hyperbola **(Minkowskian orthogonality).** In the first case it is *centripetal* (in blue) in the second case it is *centrifugal* (in red).

TANGENT $(dx/dt = \beta)$

$x': t' = t - \beta x = \tau - \beta_H x = 0$

**HYPERBOLIC ROTATION**
Uniform Hyperbolic Motion
(UHM)

**PHASE 3** CRUISE VELOCTY AND SYNCHRONOUS DISTANCE

$t_{P'} = \beta\gamma d$

**PHASE 2** ARC OF HYPERBOLA_ ACCELERATION

ACTIVE LT or BOOST
$0 \to \beta$ phase of acceleration $\alpha$

**PHASE 1** (rest)
**Center of Rotation (t=t'=0)**
$O \equiv O'$
**FOR ANY β**
1) Velocity of O' is transformed $0 \to \beta$ **instantaneously**
2) emission of photon that determines the synchronous distance d (Fig 4) X= d

$d = \alpha^{-1}$   P' (β= 0)

EUCLIDEAN ROTATION
'Uniform Circular Motion (UCM)

Arc of hyperbolic rotation
$x^2 - t^2 = \alpha^{-2} = d^2$   $(0 \le t \le \beta\gamma d)$   $d \to \gamma d$

**To be or not to be zero for α?** Relativistic acceleration α is a spacelike 4-vector whose norm is necessarily positive.

**FIG 4**
**Minkowski's proper time and Poincaré's proper length**
The reciprocal (Poincaré 1908) of process of synchronization of clocks (Einstein 1905)
"Two ways" or "round trip" light travel for distance d. (FACTOR $\gamma$)
**"One way" light travel for distance horizon $R_H$**
For one way distance $r < R_H$ we have the FACTOR k of BONDI

t

t'

EINSTEIN AXIS

2 D two ways

x' (t'= $\tau$)

M'

$\gamma$D

S

ONE WAY

D

x' (t'=0)
POINCARE AXIS
(reciprocal)

P' ($\gamma$d, $\beta\gamma$d)

d

$O \equiv O'$
EMISSION OF
RADIATION IS
USED TO
DETERMINE
DISTANT in K'

d

$\gamma$d

x

# FIG 5
## FUNDAMENTAL "LORENTZ BOOST or HYPERBOLIC ROTATION" AND COSMOLOGICAL RELATIVITY (CR)

CsR has two invariants (c and $\Lambda = \alpha^2_M$). The Radius of Hubble $R_H$ must be define by **one-way** light travel time (life time of Hubble $T_H$). Any distance **r** is defined by $R_H$, the unity (metric) of the Universe: $\beta = H\,r$ (the law of Hubble, $H = \alpha_M$, is the constant of Hubble)

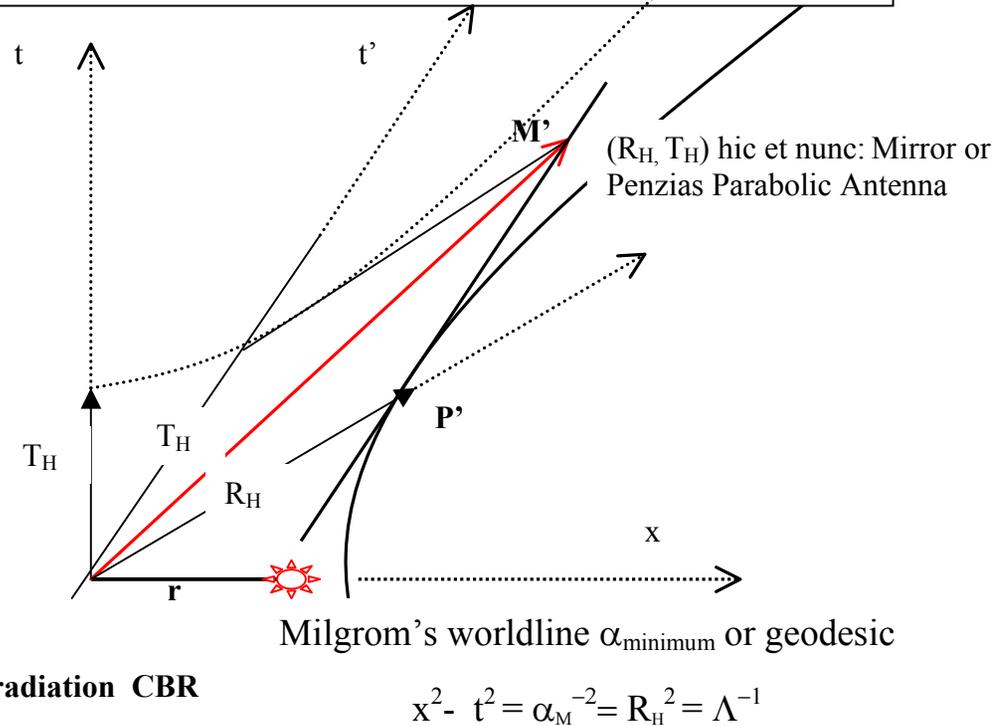

In CR, the scale factor is **Bondi's factor** k and the Hubble constant is defined, exactly as in Friedman's model, by the logarithmic derivation of this expansion factor $\ln k = \beta_H$

$$H = k'(\tau) / k(\tau)$$

**We represent remote galaxy by a "red star" (Bondi's factor is Doppler's redshift factor)**

$(R_H, T_H)$ hic et nunc: Mirror or Penzias Parabolic Antenna

$O \equiv O'$
**Emission of radiation CBR**

Milgrom's worldline $\alpha_{minimum}$ or geodesic

$$x^2 - t^2 = \alpha_M^{-2} = R_H^2 = \Lambda^{-1}$$

where $\Lambda$ is the cosmological constant
This equation is "Lorentz invariant" for any observer in uniform motion in space-time vacuum with a global curvature

**Note:** "NOT TO BE" in the light cone is the characteristic of the other (doted) along Ot hyperbola. Unlike Milgrom's word-line, which is integrated from an element of proper time $d\tau$, this hyperbola is integrated from an element of proper length $d\xi$




\begin{thebibliography}{W. Rindler \& M. Ishak (2007)}
\bibitem[H. Minkowski 1908]{1}  ''Raum und Zeit'', Cologne, Phys. Z., 20,
1909, 104-111. Teubner, Leipzig - Berlin, 5\`{e}me \'{e}dition, 1923, 54 -
71.

\bibitem[W. Rindler (1966)]{2a}  ''Kruskal space and the uniformly
Accelerated frame'', Am. J.of Physics, Vol. 34, Issue 12, December 1966,
1174-1178.

\bibitem[W. Rindler (SR) ]{2b}  ''Relativity. Special, General and
cosmological'', Oxford University Press, 2001.

\bibitem[W. Rindler (GR)]{2c}  ''Introduction to special relativity'',
Oxford Science Publication.

\bibitem[W. Rindler \& M. Ishak (2007)]{2d}  ''The Contribution of the
Cosmological Constant to the Relativistic Bending of Light Revisited'',
Phys. Rev. D 76, 043006 (2007) [5 pages]

\bibitem[M. Milgrom]{3a}  ''A Modification of the Newtonian Dynamics as a
Possible Alternative to the Hidden Mass'', Astrophys J., 270, 365-370, 1983.

\bibitem[M. Milgrom MOND]{3b}  ''The MOND paradigme'',
arXiv:08013133v2[astro-ph] 3 Mars 2008.

\bibitem[J. D. Anderson and all]{4}  ''Study of the anomalous acceleration
of Pioneer 10 and 11''. Phys. Rev. D. 65, 082004 (2002)

\bibitem[L. Blanchet]{5}  ''Dipolar Dark matter and MOND'', \ s\'{e}minaire
\`{a} l'ULB, invit\'{e} le 23 f\'{e}vrier 2009 par B. Famaey.

\bibitem[E. Gunzig]{6}  Que faisiez-vous avant le big bang? Odile Jacob,
2008, Paris.

\bibitem[A. Einstein (1905)]{7}  \ ''Zur Elektrodynamik bewegter
K\"{o}rper'', Ann.d.Ph,17, p892-921,1905.

\bibitem[M. Born ]{8}  ''Die theorie des Starren Elekctrons in Kinematics
des Relativit\"{a}tPrinzip'', Ann. Phys, Band 30 (1909), 1-56.

\bibitem[G.\ Ellis \& R.\ Williams]{9}  ''Flat and curved space-times'',
Oxford University Press, 2000.

\bibitem[H. Poincar\'{e} 1908]{10}  ''La dynamique de l'\'{e}lectron''. R.
G\'{e}n. des Sciences Pures et Appl., 19, 386-402, 1908.

\bibitem[Y. Pierseaux 2008]{11}  ''La cin\'{e}matique sous-jacente \`{a}
l'ellipse de Poincar\'{e}'', C.R. Physique, n$%
{{}^\circ}%
$ 7-8, p921-929, 2007, Paris

\bibitem[Y. Pierseaux 2004]{11a}  ''Einstein's spherical light waves versus
Poincar\'{e}'s ellipsoidal light waves'', An. F. de Broglie, vol 30, n$%
{{}^\circ}%
3-4,$ 2005.

\bibitem[Y. Pierseaux (2009)]{11b}  ''Du boost de Lorentz \`{a}
l'acc\'{e}l\'{e}ration de Milgrom'', IIHE, 2009-01, ULB-VUB, Pleinlaan, 2,
1050 Brussels.

\bibitem[Y. Pierseaux 1999]{12}  ''La structure fine de la Relativit\'{e}
Restreinte, L'Harmattan, Paris, 1999.

\bibitem[H. Bondi ]{13}  ''Assumption and Myth in Physical Theory'',
Cambridge University Press, 1967.

\bibitem[R. Penrose]{14}  ''The large, the small and the human mind'',
Cambridge University press, 2000.
\end{thebibliography}
\end{document}